\documentclass[preprint2]{aastex}
\shorttitle{Sco X-1}
\shortauthors{Fomalont, Geldzahler \& Bradshaw}
\begin{document}

\title{Sco X-1: Energy Transfer from the Core to the Radio Lobes}

\author{E.~B.~Fomalont}
\affil{National Radio Astronomy Observatory, Charlottesville, VA 22903}
\email{efomalon@nrao.edu}

\author{B.~J.~Geldzahler \& C.~F.~Bradshaw}
\affil{School of Computational Sciences, George Mason University, Fairfax, VA 22030}
\email{bgeldzahler@hotmail.com \& cbradshaw@tstag.com}

\begin{abstract}
    The evolution of the radio emission from Sco X-1 is determined
from a 56-hour continuous VLBI observation and from shorter
observations over a four-year period.  The radio source consists of a
variable core near the binary, and two variable compact radio lobes
which form near the core, move diametrically outward, then fade away.
Subsequently, a new lobe-pair form near the core and the behavior
repeats.  The differences in the radio properties of the two lobes are
consistent with the delay and Doppler-boosting associated with an
average space velocity of 0.45c at $44^\circ$ to the line of sight.
Four lobe speeds, between 0.32c to 0.57c, were measured for several
lobe-pairs on different days.  The speed during each epoch remained
constant over many hours.  The direction of motion of the lobes over
all epochs remained constant to a few degrees.

     Two core flares are contemporaneous with two lobe flares after
removal of the delay associated with an energy burst moving with speed
$\beta_j>0.95$ in a twin-beam from the core to each lobe.  This is the
first direct measurement of the speed of energy flow within an
astrophysical jet.  The similarity of the core and lobe flares
suggests that the twin-beam flow is symmetric and that the core is
located near the base of the beam.

\end{abstract}


\section{Introduction}

    Sco X-1 is a low-mass binary system (LMXB) with strong, persistent
X-ray emission.  It is a \lq Z-type\rq~LMXB, named for a
characteristic shape of its X-ray color-color diagram.  It also
exhibits quasi-periodic X-ray oscillations (Hasinger \& van der Klis
1989; van der Klis et al.\ 1996).  The binary has an orbital period of
0.787d \citep{got75}, the degenerate object is probably a neutron
star, with a one solar mass companion of unknown spectral type (Cowley
\& Crampton 1975).

    From eight Very Long Baseline Array (VLBA) observations between
1995 and 1998, a distance of $2.8\pm 0.3$ kpc was obtained from
measurements of the trigonometric parallax of Sco X-1 \citep{bra99}.
The radio emission also varied over tens of minutes and relativistic
motion was detected.  Thus, Sco X-1 exhibited the properties of
Galactic-jet sources \citep{mir99} with similarities to extragalactic
radio sources (Blandford \& Rees 1974).  However, these observations
were too short and too intermittent to determine the properties of Sco
X-1.

     We, therefore, observed Sco X-1 with milliarcsecond (mas)
resolution over a 56-hour period on 1999 June 11-13 (MJD 51340-2).
Extensive optical and X-ray observations were also made.  This letter
reports on the general properties of Sco X-1 with emphasis on the
kinetic properties of the source.  Further discussions of the radio
observations and the nature of Sco X-1 are given in Paper II
\citep{fom01}.  The nature of the x-ray emission and accretion
processes in Sco X-1, and the interaction of the radio core and the
binary system are given elsewhere \citep{tit01,bra02, gel02}.

\section{The Radio Observations:}

    The observations in 1999 June consisted of seven consecutive
8-hour VLBI observations cycling among three different arrays: the
VLBA with the VLA, the APT (Asia-Pacific Telescope: Australia
Telescope Compact Array, Ceduna, Hartebeesthoek, Kashima, Mopra,
Parkes and Shanghai), and the EVN (European VLBI Network: Effelsberg,
Jodrell Bank, Medicina, Noto, Westerbork, plus Hartebeesthoek and the
Green Bank 140-ft).  With the VLBA, 1.7 GHz and 5.0 GHz observations
were interleaved in time.  With the APT and EVN, observations were
made at 5.0 GHz only.  All data associated with the observations of
Sco X-1 were calibrated using simultaneous observations of a nearby
background 10-mJy radio source\footnote[1]{This source is one of the
two radio objects which were previously thought to be associated with
Sco X-1.  They are unrelated background objects \citep{fom91} at a
distance from Sco X-1 of $70''$ in position angle $30^\circ$. The
radio emission from Sco X-1 extends $<0.1''$ in position angle
$54^\circ$.}, by which all images were registered to an accuracy
of $<0.1$ mas \citep{bra99}.  Because of the significant flux density
variations and component motions, images were made from data sets as
short as 50 minutes (snap-shots) in order to follow the changes in the
radio emission.

\section {The Radio Snap-shots of Sco X-1}

    The sequence of the radio snap-shots of Sco X-1 during the 1999
June observations is displayed in Figure 1.  More frequent,
higher-resolution 5-GHz snap-shots are displayed in Paper II.  The
radio structure showed three components: a radio core within a few mas
of the binary; a compact moving component north-east of the core,
called the NE lobe, and a fainter component south-west of the core,
called the SW lobe, detected most strongly on MJD 51342.
Occasionally, faint extended emission was detected between the core
and NE lobe.

    The flux density variations of the three major components are
shown in Figure 2.  Sco X-1 evolved through several stages during
these observations.  The first NE lobe disappeared at MJD 51340.4
after reaching a distance of 20 mas from the core.  The core
component, on the other hand, increased in flux density during this
period.  For the next nine hours, Sco X-1 consisted of only the core
component.  At MJD 51340.7 the core brightened and by MJD 51340.9 a
{\it new} NE lobe (\#2) emerged from the core.  Some extended emission
between these two separating components persisted until MJD 51341.3,
during which time the flux density of both components decreased.  Over
the next 30 hours the NE lobe moved directly away from the core,
reaching a separation of 50 mas.  Both the core and NE lobe varied
significantly in flux density; The NE lobe peak (N3) at MJD 51341.1
and the core peak (C4) at MJD 51341.9 reached 20 mJy at 5 GHz.  There
was no apparent correlation of the core flares with the phase of the
binary orbit.  The SW lobe, detected most strongly on MJD 51342, was
located about half as far from the core as the NE lobe.

   For the observations between 1995 and 1998, Sco X-1 was relatively
weak and an accurate motion of the NE lobe could not be measured.  The
one exception was the observations on 1998 February 27-28 1998 (MJD
50871-2), consisting of two six-hour VLBA observations separated by 18
hours, and these images are shown in Paper II, Figure 5.  On MJD
50871, the NE lobe was about 20 mas from the core, and moved away at a
speed similar to that in 1999 June.  The SW lobe was also present and
its motion, in the opposite direction as the NE lobe, was measured.

\section {Lobe Velocities and Relativistic Effects}

   Figure 3 shows the observed separation of the NE and SW lobes from
the core.  For the 1999 June observations the velocity in the plane of
the sky of the NE lobe (\#1) was $v=0.73\pm 0.07$ mas hr$^{-1}$. (For
a measured distance to Sco X-1 of 2.8 kpc, one mas = 2.8 AU =
$4.19\times 10^{8}$ km.  A projected velocity of 1 mas hr$^{-1}$ =
0.388c.)  After emerging from the core, the NE lobe (\#2) lobe speed
was $v=1.74\pm 0.16$ mas hr$^{-1}$.  At MJD 51341.4 the NE lobe flared
and its speed abruptly decreased to $v=1.25\pm 0.05$ mas hr$^{-1}$.
Although the NE lobe varied considerably in flux density over the next
30 hours, its speed remained nearly constant.  For the 1998 February
observations, the velocity of the NE lobe on MJD 50871 was $v=1.11\pm
0.06$ mas hr$^{-1}$, with no significant departure from uniform
motion.

    The speeds for the weaker SW lobes were determined less
accurately.  On MJD 50872 $v=0.6\pm 0.2$ mas hr$^{-1}$, and on MJD
51342 $v=0.5\pm 0.3$ mas hr$^{-1}$.  Further evidence of the motion of
the SW lobe is described in Paper II which shows that whenever the SW
lobe is detected, its distance from the core is always about half that
of the NE lobe.  Since the NE lobe was clearly moving away from the
core, this ratio is also equal the ratio of the speed of the SW lobe,
$\beta_{SW}$, to the speed of the NE lobe, $\beta_{NE}$ and is
$\beta_{SW}/\beta_{NE}=0.51\pm 0.02$ using all observations of Sco
X-1.  Such lobe velocity differences have been observed in other
Galactic-jet sources \citep{mir99} and are caused by delay effects
from their relativistic motion.

     The expected Doppler-boosting (attenuation) of the approaching
(receding) component was observed in Sco X-1.  An analysis in Paper II
derives an average flux density ratio between the SW and NE lobes of
$0.12\pm 0.03$.  On the other hand the spectral index, $\alpha$
(S$\propto \nu^\alpha$), of the two lobes should not be affected
significantly by a large space motion and, indeed, both lobes had
$\alpha\approx -0.6$ during the flaring states.  The lobe angular
diameters were also similar.  This comparison of the radio emission
from the NE and SW lobes suggests that they are intrinsically similar,
but their appearances are strongly affected by the delay and Doppler
effects of their space motion.

     With an accurate distance to Sco X-1, the lobe space motion can
be determined solely from the observed lobe kinetics.  Although four
speeds between 0.28c and 0.68c {\it in the plane of the sky} were
measured for the NE lobe, there were two periods when the lobe speed
remained constant for over one day: MJD 50871-2 with
$\beta_{NE}=0.43\hbox{c}\pm 0.02\hbox{c}$ and MJD 51341-2 with
$\beta_{NE}=0.49\hbox{c}\pm 0.02\hbox{c}$.  Thus, we adopt a typical
speed of the NE lobe of $\beta_{NE}=0.46\hbox{c}\pm 0.08\hbox{c}$ with
the estimated error derived from the spread of the measured speeds and
the distance uncertainty to Sco X-1.  Since the position angle of the
radio axis of Sco X-1 in the plane of the sky over the five year
observation period remained within a few degrees of $54^\circ$ (Paper
II), the direction of the space velocity of the lobes was also
constant to this same level.  Hence, any change in the observed speed
of the lobes is caused by changes in speed and not in direction.
With the derived values of $\beta_{NE}$ and $\beta_{SW}/\beta_{NE}$,
the average space velocity of the lobes can be determined
\citep{bla77}:
\begin {equation}
\frac{\beta_{SW}}{\beta_{NE}} =0.51\pm 0.02 = \frac{1-\beta~\hbox{cos}(\theta)}{1+\beta~\hbox{cos}(\theta)}
\end {equation}
\begin {equation}
\beta_{NE} = 0.46\pm 0.08 = \frac{\beta~\hbox{sin}(\theta)} {1-\beta~\hbox{cos}(\theta)}
\end{equation}
where $\beta$ is the true speed of the lobes moving in direction
$\theta$ with respect to the observer.  The average {\it space
velocity} of the lobes is then $\beta = 0.45\pm 0.03;~\theta =
44^\circ \pm 6^\circ$.  The quoted errors are the one-sigma
uncertainty.  The four measured space velocities for different
lobe-pairs were: 0.32c, 0.43c, 0.46c and 0.57c.

    The flux density ratio for the SW to NE lobe from Doppler-boosting
expected from the derived space velocity is
$R=(\beta_{SW}/\beta_{NE})^{k-\alpha}$ \citep{bla79}.  For k=2, which
is appropriate for an optically thin discrete component, and
$\alpha=-0.6$ as measured, we predict R$=(0.51)^{-2.6}\approx 0.17$.
However, if the magnetic field distribution in the lobes is
anisotropic, the flux ratio could be considerably smaller
\citep{caw91}.  The observed flux density ratio, R=0.12 is in
reasonable agreement with that expected from the Doppler boosting
which was derived {\it solely} from the lobe motions.

\section {The Correlation of Flux Density Variations and the Jet
Speed}

    The large flux density variations of the core and the lobes during
the 1999 June observations are shown in Figure 2.  The flare
characteristics of all components were similar; a factor of 2 to 10
increase in flux density; $-0.3<\alpha<-0.6$ (except for flare C1); a
sharper rise time than decay time; a flare width of about three hours;
and a lobe angular size $<3$ mas.

    In Paper II we suggest that the NE and SW lobes are moving
hot-spots whose internal and kinetic energies are supplied by the
disruption of the energy flow in an oppositely-directed twin-beam
formed near an accretion disk around the neutron star.  Therefore, we
searched for a correlation of the intensity variations of the lobes
versus the core using the following model: (1) A core flare responds
in an unspecified manner to an event near the binary system; (2) This
event is also associated with an energy surge which propagates down
the twin-beam with an average velocity of $\beta_{j}$; (3) Each lobe
flares when this increased energy flux reaches them.  With this model
the expected delay, {\it as seen by an observer}, between a core flare
and its manifestation in the NE and SW lobes can be determined since
the lobe space velocity, $\beta$ and $\theta$, are accurately known.
The expected delays, $\tau_{NE}, \tau_{SW}$, are:
\begin {equation}
\tau_{NE} = (t_1-t_0)\beta \frac{(1-\beta_j\hbox{cos}\theta)}{(\beta_j-\beta)}
\end {equation}
\begin {equation}
\tau_{SW} = (t_1-t_0)\beta \frac{(1+\beta_j\hbox{cos}\theta)}{(\beta_j-\beta)}
\end {equation}
where $t_0$ is the time of ejection of the lobes from the core, and
$t_1$ is the time of a core flare.  The $(\beta_j-\beta)^{-1}$ factor
is related to the time for the flare \lq information\rq~to travel in
the beam from the core to the lobes.

    Figure 4 shows the flux density variations of the three
components, after removal of the delay associated with the NE and SW
lobes for $\beta_j=1.0$ and near the flare peaks for $\beta_j=0.9$.
The temporal correlations for the flares C3-N3 ($\tau_{NE}=2.8$h),
C4-N4 ($\tau_{NE}=7.4$h) and C3-S3 ($\tau_{SW}=19.7$h) with
$\beta_j=1.0$ are excellent, and the sensitivity of these correlations
to $\beta_j$ are illustrated from those points plotted for
$\beta_j=0.9$.  A fit of the correlation of the three observed
flare-pairs as a function of the jet speed gives a formal solution,
$\beta_j=1.02\pm 0.04$ (Paper II).  Even with these limited statistics
and an over-simplified model, these observations suggest that the
speed of energy flow in the beam is $>0.95$c.  This result is the
first direct measurement of energy flow velocity in an astrophysical
beam.  The velocity close to the speed of light suggests that the
energy flux is probably carried by leptons with $\gamma>3$, rather
than with protons or heavier particles moving at velocities
considerably less than c \cite{bla74}.

    For flare 3, the Doppler-corrected flux densities, the general
shapes, and spectral properties of the NE and SW lobes are similar.
This suggests that the energy flow in the twin-beam is symmetric, and
that the production of radio energy in each lobe depends mainly on the
beam energy flow, rather than on interaction details at the working
surfaces in the lobes.  For flare 4 the flux density of the core is
about ten greater than that for the NE lobe.  (This flare in the SW
lobe did not occur until thirty hours after the termination of the
experiment.)  Possible explanations on the relative weakness of the NE
lobe for this flare are discussed in Paper II.
 
    The similar properties of the two core flares with the lobe flares
are surprising since the processes producing the radio emission in the
lobes and in the core are thought to be different.  Since these core
flares are clearly associated with the energy flow in the beam, we
suggest that the core component is located near the base of the beam
rather than within the binary system.  A symmetric \lq
core\rq~component at the base of the beam to the south-west of the
binary should also be present, but would be strongly
Doppler-attenuated.  Further discussion of the relationship between
the radio core and the binary system are given elsewhere
\citep{gel02}, and includes a discussion of flare 2 which is
associated with the formation of a new pair of lobes, and flare 1
which is anti-correlated with NE lobe in the sense that the NE lobe
vanishes when the core reached a maximum.

\section {Conclusions}

    The relatively simple and recurrent nature of the radio properties
of Sco X-1 permit the determination of its kinetics.  The compact
lobes move directly away from the core with an average velocity of
$0.45c\pm 0.03c$ at an angle $44^\circ\pm 6^\circ$ to the line of
sight, with a range for four different lobe-pairs between 0.32c to
0.57c.  The lobe advance remains constant for many hours despite
significant flux density variations.  The excellent correlations among
some of the core and lobe flares suggest that energy flux within the
twin-beam flows at a speed $>0.95$c, that the energy flow is symmetric
in the twin-beams, that the radio luminosity of the lobes and core
depends strongly on the energy flux in the beam but weakly on other
environmental parameters, and that the location of the radio core is
more likely to be at the base of the beam rather than near the binary
system or accretion disk.

\acknowledgements

The National Radio Astronomy Observatory is a facility of the National
Science Foundation, operated under cooperative agreement by Associated
Universities, Inc.  We thank the European VLBI Network (EVN) and the
Asia-Pacific Telescope (APT) for their support and observation time.
The data were correlated with the VLBA correlator in Socorro and the
Penticton Correlator which is supported by the Canadian Space Agency.
It is a pleasure to thank Dr.\ Jean Swank for granting us RXTE time,
to Dr.\ Tasso Tzioumis for help with the APT scheduling, to Dr.\ Sean
Dougherty for processing the APT data.  We also thank Dr. Michael
McCollough for comments on the draft.

\begin {figure*}
\vskip -0.0in
\hskip -0.6in
\epsscale{1.80}
\plotone{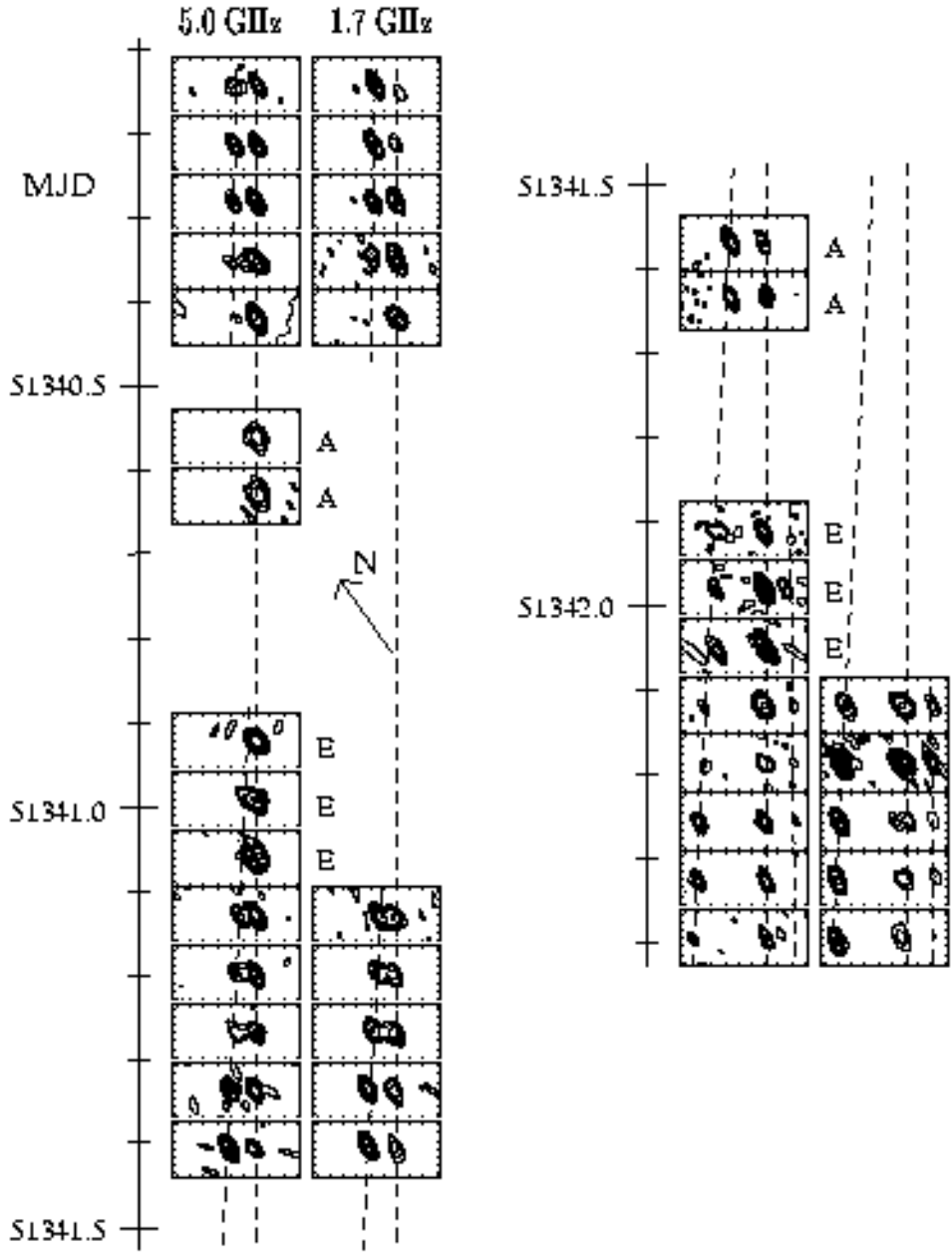}
\vskip 0.0in
\caption[fig1_let.ps] {{\bf Snap-shots of Sco X-1 with $10\times 5$
mas resolution during the 1999 June observations.} The 5.0 GHz and 1.7
GHz snapshots are shown side by side with the time axis on the left of
the two vertical blocks.  Each image has been rotated $36^\circ$ with
respect to North, indicated by the arrow near the center of the
figure.  The contour levels at 5.0 GHz are 0.5 mJy
$\times~(-1,1,2,4,8\ldots)$, and at 1.7 GHz are 0.7 mJy
$\times~(-1,1,2,4,8\ldots)$.  The tick marks are separated by 10 mas
along the abscissa and 5 mas along the ordinate.  Snapshots labeled
with A are from the APT observations, those with E are from the EVN
observations.  All others snapshots are from the VLBA.  The vertical
dashed-line indicates the location of the core (determined from
previous radio observations), the line to the left the location of the
NE lobe (\#1) and the NE lobe (\#2).  The line to the right, for MJD
51342 only, shows the location of the SW lobe.}
\end{figure*}

\begin{figure*}
\vskip -3.0in
\hskip -2.0in
\epsscale{2.5}
\plotone{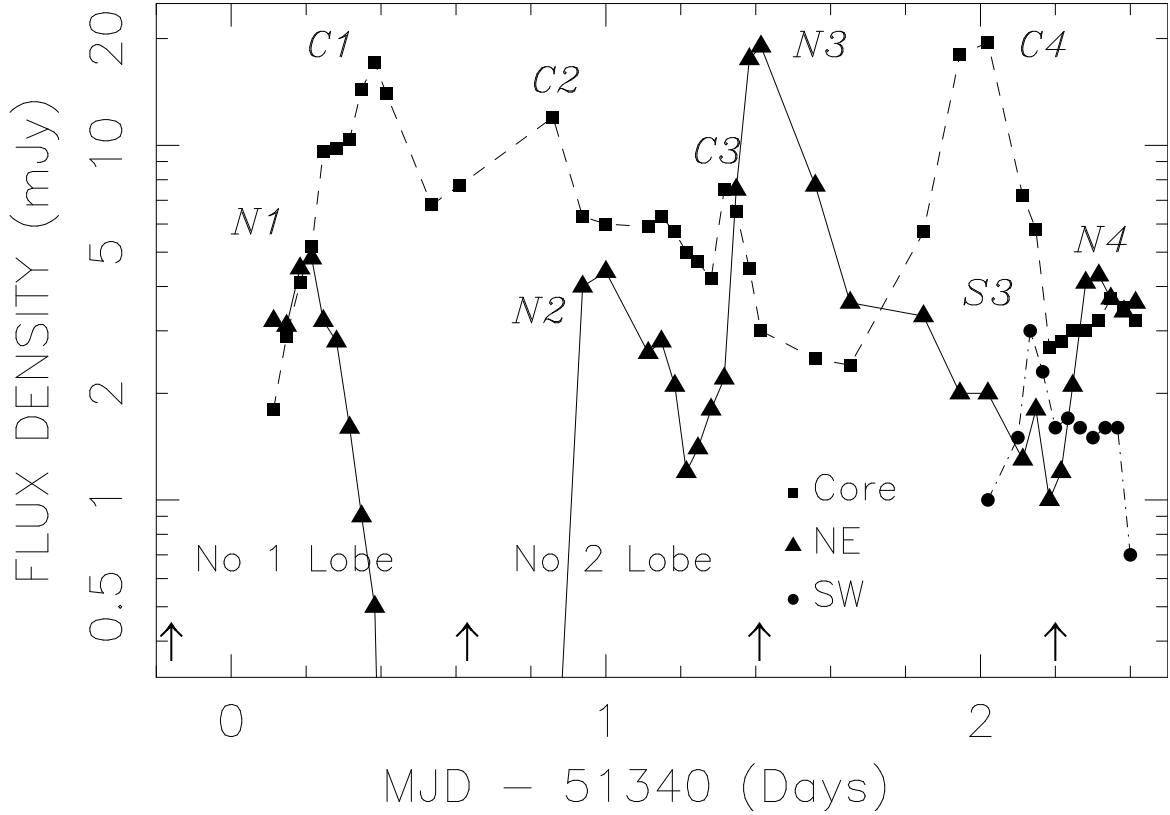}
\vskip -2.0in

\caption [fig2_let.ps] {{\bf The Flux Density of the Components in
Sco X-1 during the 1999 June Observations.} The 5-GHz flux density for
the core and NE lobe, and the 1.7-GHz for the SW lobe are plotted.
(The weak SW lobe was detected more often at 1.7 GHz than at 5 GHz.
Its flux density at 5 GHz is about half that at 1.7 GHz.) A typical
error is 0.2 mJy. A flux density determination was made every 50
minutes for the VLBA observations and every two hours for the APT and
EVN observations.  The major flares are indicated: C=core, N=NE lobe,
S=SW lobe.  The arrows near the abscissa show the times of minimum
optical light for the binary system.}
\end{figure*}

\begin{figure*}
\vskip -4.7in
\epsscale{3.7}
\hskip -4in
\plotone{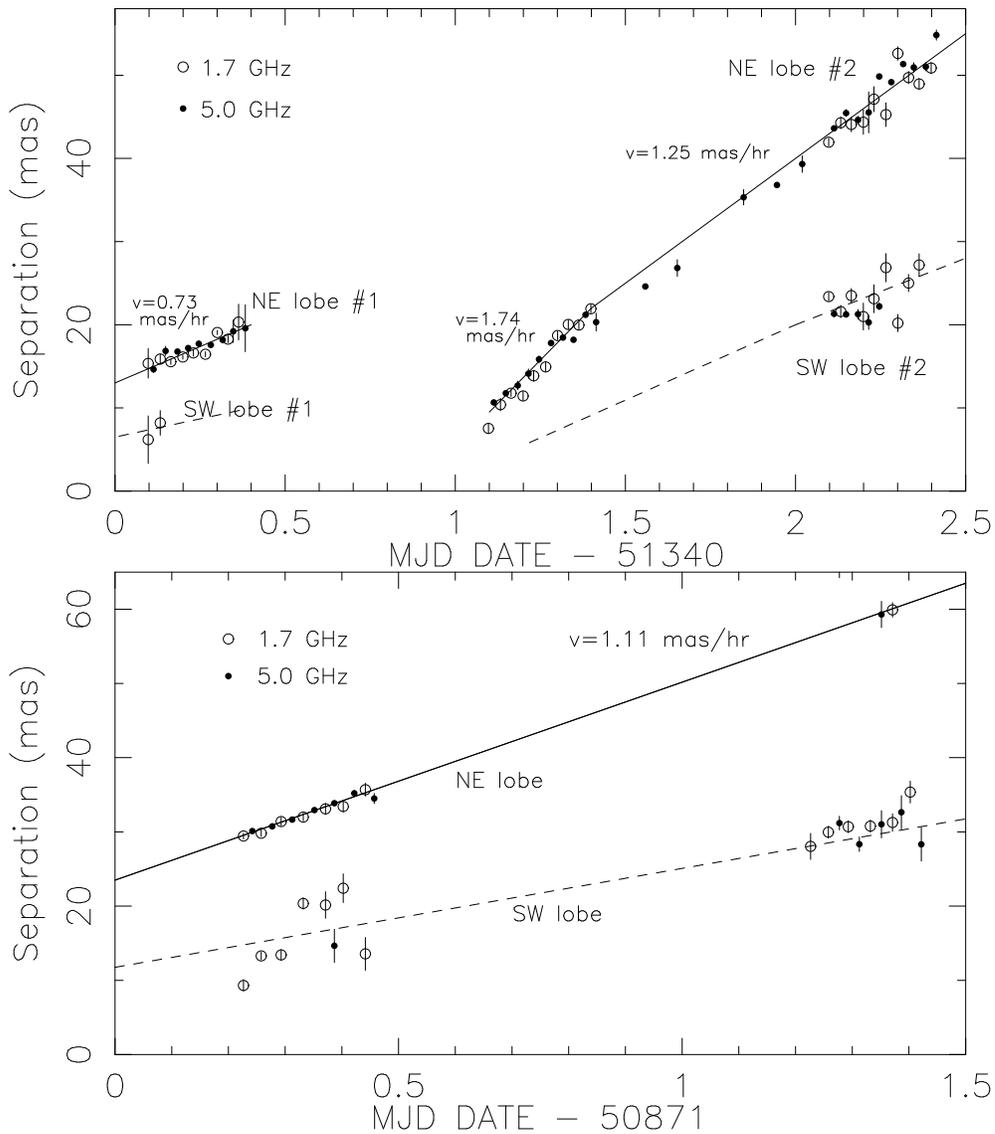}
\vskip -3.8in
\caption[fig3_let.ps] {{\bf The Motion of the NE and SW Lobes.} (top)
1999 June Observations and (bottom) 1998 February Observations.  The
plotted points show the separation the NE and SW lobes from the core
component measured every 50 min from the VLBA snaphots at 1.7 and 5.0
GHz and every two hours at 5.0 GHz from the APT and EVN for 1999 June.
Many of the indicated one-sigma error bars are smaller than the symbol
size.  The best linear velocity fits for NE lobe (\#1) and lobe (\#2) in
1999 June and for the NE lobe in 1998 February are indicated by the
solid line.  The dashed lines through the SW components are not fits
to the data, but 50\% of that of the NE lobe at the same time.}
\end{figure*}

\begin{figure*}
\vskip -3.0in
\hskip -2.0in
\epsscale{2.5}
\plotone{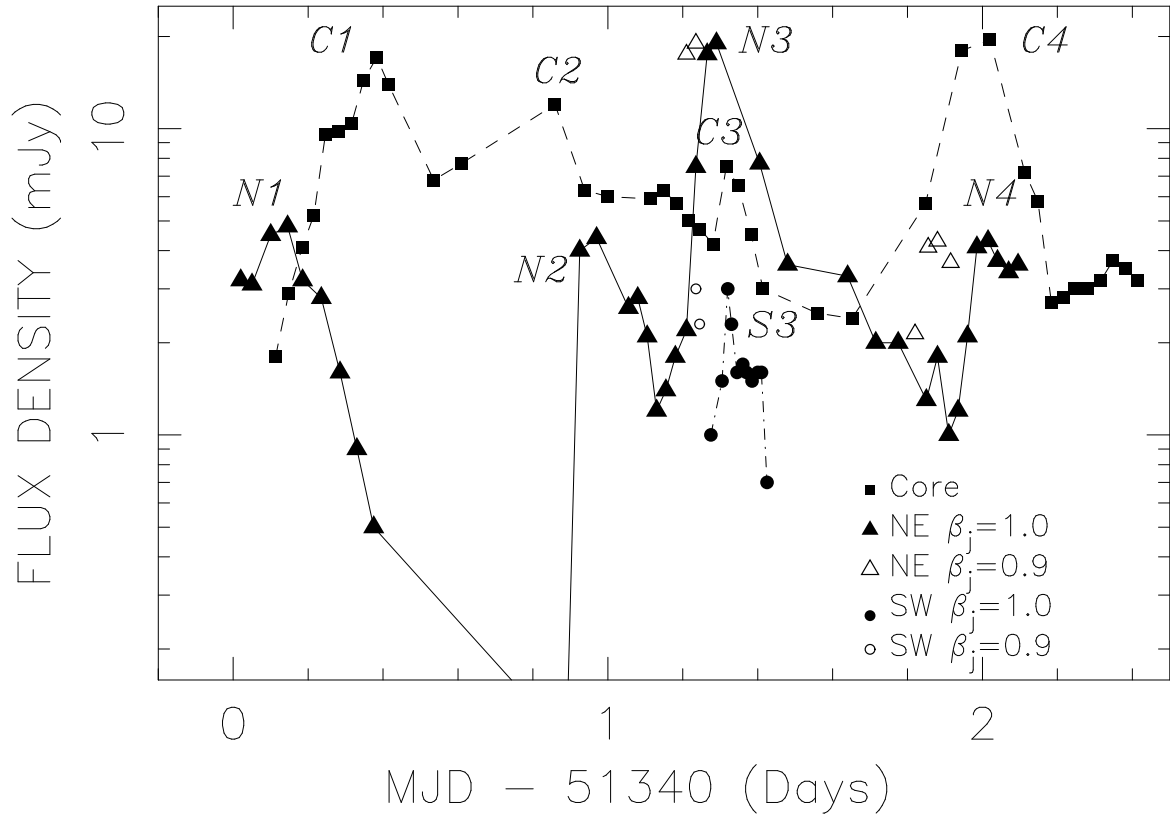}
\vskip -2.0in
\caption[fig4_let.ps] {{\bf The Flux Density of the Components in Sco
X-1 during the 1999 June Observations After Correction for the Time
Delay.} The jet speed is $\beta_j=1.0$.  This plot is similar to
Figure 2, but with the NE and SW lobe times modified as discussed in
the paper.  Several points near the N3, N4 and S4 peaks are given for
$\beta_j=0.9$.  }
\end{figure*}
\end{document}